\documentclass[12pt]{aipproc}

\usepackage{natbib}
\usepackage{graphicx}

\layoutstyle{6x9}

\begin{document}

\title{Barium Ions for Quantum Computation}
\author{M. R. Dietrich, A. Avril, R. Bowler, N. Kurz, J. S. Salacka, G. Shu and B. B. Blinov}{address={University of Washington Department of Physics, Seattle, Washington, 98195}}
\keywords{Quantum Computation, Barium, Trapped Ions}
\classification{03.67.Lx, 37.10.Ty, 42.50.Dv}
\copyrightholder{Matthew Dietrich}
\copyrightyear{2008}

\begin{abstract}
Individually trapped $^{137}$Ba$^+$ in an RF Paul trap is proposed as a qubit candidate, and its various benefits are compared to other ionic qubits.  We report the current experimental status of using this ion for quantum computation.  Future plans and prospects are discussed.
\end{abstract}

\maketitle

\section{Introduction}

At present, trapped ions are the leading contender for use as a qubit in quantum computation schemes.  This is a result of the high degree of motional control possible over the ion, the availability of long established techniques for quantum manipulation of trapped ions, and the success of shelving schemes as a highly efficient readout mechanism.  To date, Be$^+$~\cite{Turchette1998},  Ca$^+$~\cite{Blatt2004,Lucas2007},  Cd$^+$~\cite{Lee2003}, Mg$^+$~\cite{Schaetz2007}, Sr$^+$~\cite{Letchumanan2007}, and Yb$^+$~\cite{Olmschenk2007} have all been demonstrated as possible ionic qubits.  It is possible to create ionic qubits using either two hyperfine levels~\cite{Blinov2004b} or, in some species, two levels separated by an optical transition~\cite{Blatt2004} as the computational basis.  We propose to use the ground state hyperfine levels of $^{137}$Ba$^+$ as a qubit.  Although Ba$^+$ was the first ion trapped in isolation~\cite{Dehmelt1980}, $^{137}$Ba$^+$ was not trapped for another 20 years~\cite{DeVoe2002}.  The spectroscopic properties of Ba$^+$ have been carefully studied since then because of its potential applications as an optical frequency standard~\cite{Koerber2002} and in a test of parity non-conservation~\cite{Fortson1993}.  It has several desirable qubit properties, including visible wavelength transitions, high natural abundance of the $^{137}$Ba isotope, and a long lived shelving state.  Here we demonstrate single qubit initialization, rotations and readout on this new qubit, and discuss the future directions of our work.

\section{Experimental Setup}

Ba$^+$, like some of its qubit competitors such as Ca$^+$ and Sr$^+$, has an energy level structure that includes two low lying, long lived D states - see Fig. 1(a).  In contrast with these ions, however, all of the Ba$^+$ dipole transitions lie in the visible spectrum, which greatly simplifies laser alignment.  The doppler cooling consists of a blue 493 nm transition from the ground state to the $P_{1/2}$ state, which has a branching ratio of 0.244~\cite{Davidson1991} to $D_{3/2}$.  Because this state is long lived, a repump laser at 650 nm is necessary for continuous cooling.  However, the upper D state, $D_{5/2}$, is isolated from the cooling cycle and so constitutes a ``dark'' state which can be used for high fidelity readout.  Its lifetime of 35 s helps reduce the overall error rate during readout, compared to the relatively short lifetime of, for example, Ca$^+$ at 1 s.  The predominant isotope of Ba is 138 (72\% abundance), which has no nuclear spin and so no hyperfine structure to use as a qubit.  When trapping the odd isotope of Ba$^+$ (11\% abundance), all the various hyperfine levels must be addressed by introducing sidebands onto the two cooling lasers.  For the red 650 nm laser, the $D_{3/2}$ levels lie close enough~\cite{Silverans1986} that not all possible transitions need to be covered and so only three modulation frequencies are used - 614, 539, and 394 MHz.  For the blue, modulation is introduced with frequency equal to the ground state hyperfine splitting, and the carrier is set such that all ground states are excited only into the $P_{1/2}$ (F=2) manifold, see Fig. 1(b).  This is advantageous not only for the optical pumping reasons discussed below, but also because one avoids the $S_{1/2}$ (F=1) to $P_{1/2}$ (F=1) transition which is extremely weak as a result of small, destructively interfering geometric factors~\cite{Metcalf}.

The ion can be excited directly from the F=2 ground state to the $D_{5/2}$ level using a 1762 nm fiber laser (Koheras Adjustik).  While in this shelved state, the ion will not flouresce when illuminated with the cooling lasers.  A bright ion can be distinguished from a dark one with nearly perfect fidelity after a couple ms of observation time, resulting in highly reliable readout.  Because the transition is so weak (E2 transition), resonant excitation would require a carefully stabilized laser.  However, using adiabatic passage, it is possible to perform highly efficient population transfers using only a poorly stabilized laser~\cite{Wunderlich2005} .  Using the 10 mW of available power, simulation indicates that this should be possible on the ms time scale.  Because of the many magnetic sublevels of the $D_{5/2}$ state, however, it is necessary to perform some stabilization to prevent accidental excitation into an undesired state.  

\begin{figure}
\centering
\begin{minipage}{6.6cm}
\begin{center}
\includegraphics[width=6cm]{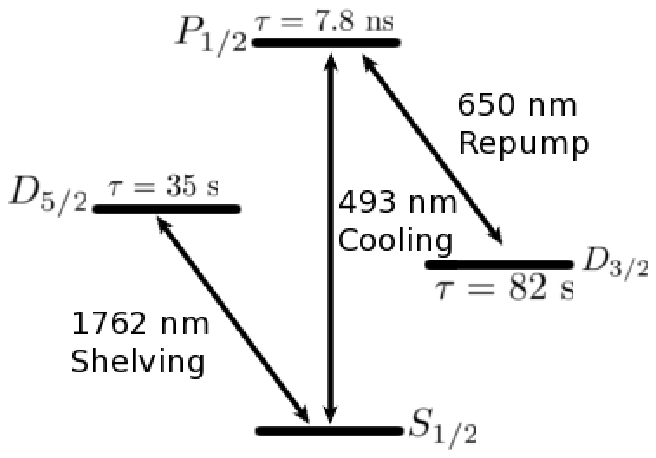}\\
(a)
\label{leveldiagram1}
\end{center}
\end{minipage}
\begin{minipage}{6.6cm}
\begin{center}
\includegraphics[width=6cm]{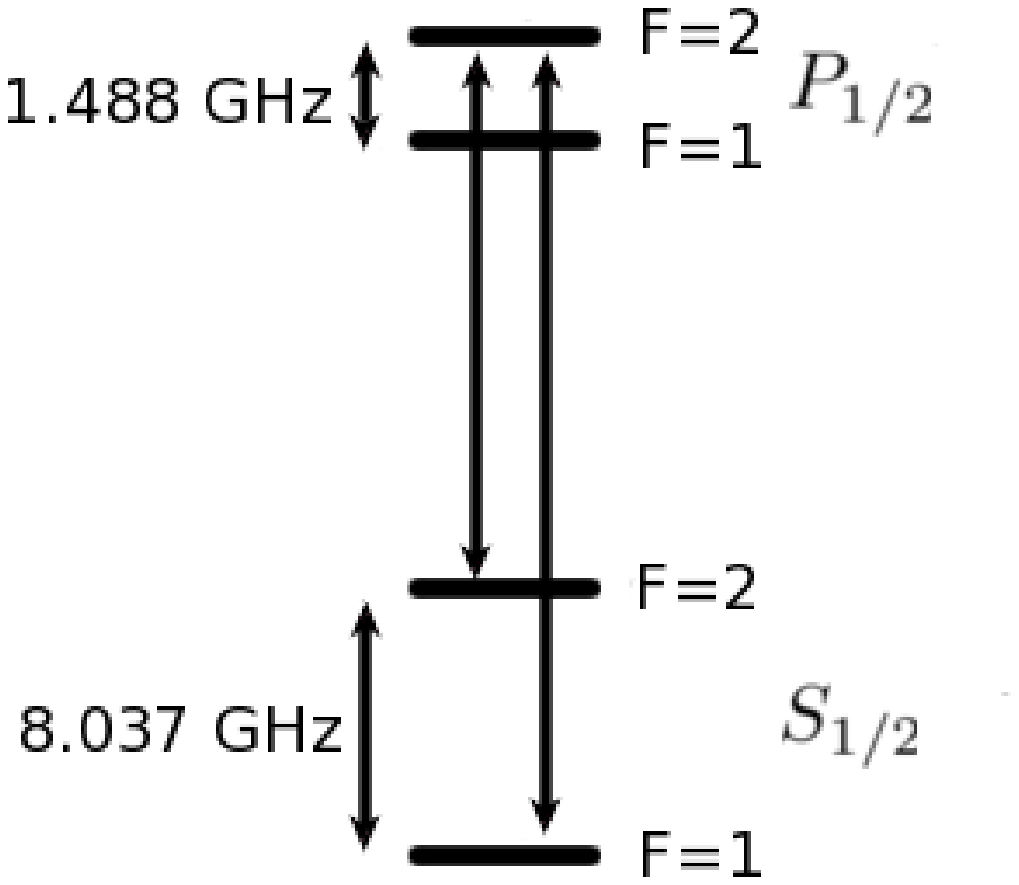}\\
(b)
\label{leveldiagram2}
\end{center}
\end{minipage}
\caption{Energy level diagram for Ba$^+$.  (a) The basic cooling and shelving scheme.  (b) A detail of the 493 cooling transition, showing the relevant transitions.  The (F=2,$\textrm{m}_F$=0) to (F=2,$\textrm{m}_F$=0) $\pi$ transition is actually forbidden, and so an elliptical polarization must be used for cooling.}
\end{figure}

The ground state of $^{137}$Ba$^+$ is split due to the hyperfine interaction by about 8.037 GHz~\cite{Blatt1981}, which makes this isotope usable as a hyperfine qubit.  We optically pump into the upper $\textrm{m}_F$=0 state with $\pi$ polarized 493 nm light, since parity symmetry forbids the $S_{1/2}$ (F=2,$\textrm{m}_F$=0) to $P_{1/2}$ (F=2,$\textrm{m}_F$=0) transition.  After state preparation, we can cause direct Rabi flops between the upper and lower magnetically insensitive $\textrm{m}_F$=0 states using microwaves at the known hyperfine frequency, exposed for varying periods of time.

\begin{figure}
\includegraphics{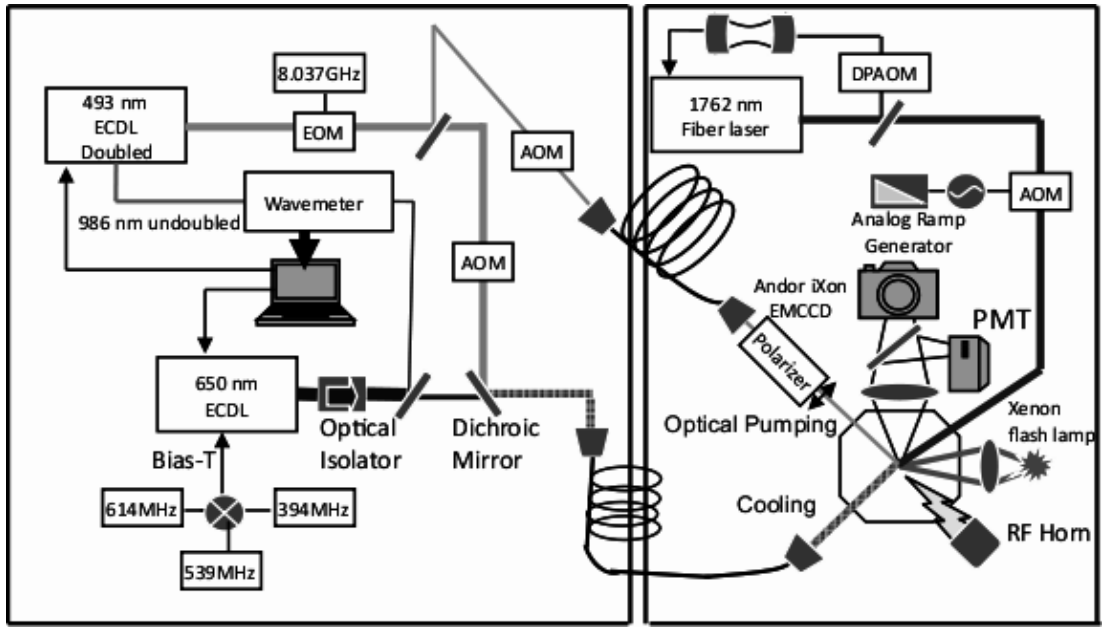}
\caption{A schematic of the experimental setup.  The cooling lasers are locked using a commercial wavemeter, which feeds back to the laser systems.  Acousto-optic modulators (AOMs) are used for high speed shuttering and, in the case of the 1762, frequency modulation and shifting (Double Pass AOM).  Barium ionization is achieved through photoionization with a Xe flash lamp.  A mirror on a motorized mount changes the detection from a CCD camera to a PMT.}
\label{apparatus}
\end{figure}

The cooling light is provided by two external cavity diode lasers (ECDL), one at 650 nm (Toptica DL-100) and the other at 493 nm (Toptica SHG-110), the latter of which is a 986 nm diode frequency doubled in a bow-tie enhancement cavity.  The available power is 10 mW and 20 mW respectively.  Both lasers are stabilized to within about 3 MHz using a high-precision wavelength meter (HighFinesse WS-7).  The blue is then modulated using a resonant EOM at 8.037 GHz (New Focus model 4851), while the red has its drive current modulated directly with a bias-T.  The two are then combined using a dichroic mirror and coupled into a single mode fiber, which provides colinearization and mode cleaning.  The fiber also allows us to send the cooling laser light into another room where the active ion trap presently resides.  A small fraction, about 10 $\mu$W, of 493 nm light is split off after the EOM for the optical pumping.  The 1762 nm fiber laser is output onto the second table, where it is stabilized using a high finesse Zerodur cavity with 500 MHz free spectral range suspended in a vacuum chamber which is temperature stabilized to within about 10 mK.  The transmitted laser intensity is monitored to maintain the lock.  The adiabatic passage is achieved by driving an AOM with a linear analog ramp, ensuring maximum adiabaticity.  To load Ba into the trap, we heat a sample of metallic barium contained in an alumina cylinder to several hundred degree Celsius  to create an atomic beam and then use a Xe flash lamp to photoionize.  The abundance of $^{137}$Ba is sufficiently high that one or two ions can be reliably trapped without the need for isotope selective photoionization. 

\section{Current Status}

\begin{figure}
\includegraphics[angle=270]{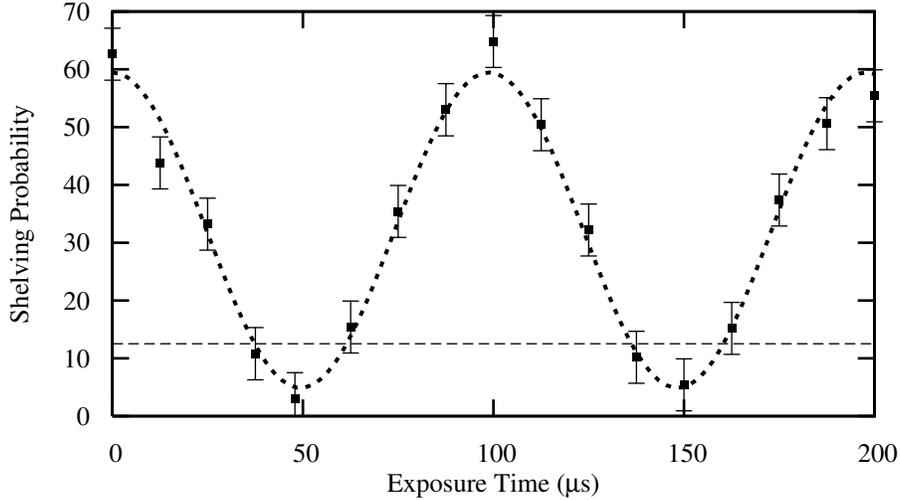}
\caption{Probability of shelving as a function of microwave exposure time.  The Rabi frequency here is 10.1 kHz.  The maximum shelving probability is limited by the stability of the shelving laser, and the minimum by optical pumping efficiency and microwave frequency detuning.  The minimum shelving observed was 4.8\%, and the contrast is 54\%.  $\chi^2_\nu=.55$ for this fit, with $\nu=13$.  The error bars are statistical.  The dashed line indicates the 12.5\% minimum which would be observed in the absence of optical pumping.}
\label{rabiflops}
\end{figure}

After qubit state preparation by optical pumping we drive Rabi flops by applying microwaves and detect the final state with the 1762 nm laser as described above.  The resulting detected Rabi flops are shown in Fig. \ref{rabiflops}.  At the time of publication, the 1762 nm laser stabilization was not complete, so that the 1762 frequency sweep crossed several transitions, resulting in only 60\% efficient population transfer from the $S_{1/2}$ (F=2, $\textrm{m}_F$=0) to the $D_{5/2}$ (F=3) and (F=4) manifold.  It should be noted that without optical pumping the lowest possible shelving probability would be 12.5\% since all ground state $\textrm{m}_F$ levels would be populated evenly, and so this data simultaneously illustrates optical pumping.

We have also recently demonstrated a full optical Rabi flop on a single $^{138}$Ba$^+$ using a single ultrafast laser pulse of 400 fs duration on the $S_{1/2}$ to $P_{3/2}$ transition at 455 nm~\cite{Kurz2008}.  This allowed the measurement of the branching ratios of the $P_{3/2}$ state with high precision.  Branching ratio measurements provide an experimental test of computational models of Ba$^+$ atomic structure, which are important for the parity non-conservation test~\cite{Fortson1993}.  Also, when compared against astronomical measurements of the branching ratio, it provides a bound on prehistoric variations in $\alpha$.   The ultrafast excitation of single Ba ions paves the way to the ion-photon~\cite{Blinov2004} and remote ion-ion~\cite{Moehring2007} entanglement.

\section{Future Plans}

If tuned to the $P_{1/2}$ transition, ultrafast Rabi flops could be made to apply a state dependent impulse on the ion, which would allow us to perform ultrafast gates, such as the Garc\'ia-Ripoll phase gate~\cite{GarciaRipoll2003}, on two ions in the same trap.  The benefits of such a gate are its intrinsic speed and that it does not necessitate cooling to the ground state of motion.  This, combined with a laser for stimulated Raman transitions between the hyperfine levels, will form the basis of quantum computing with barium.

A more immediate objective with Ba$^+$ is the remote entanglement of two ions seperated by a kilometer or more.  Remote entanglement of ions is accomplished by first exciting each ion with the ultrafast laser and allowing it to spontaneously decay, resulting in a photon whose frequency or polarization state is entangled with the final spin state of the relaxed ion.  Once this is done with two separate ions, those photons can be jointly measured in the appropriate parity basis, resulting in an entangled state between the two ions.  Entanglement between a photon and ion was originally seen in Cd$^+$~\cite{Blinov2004} and between two ions just recently in Yb$^+$~\cite{Moehring2007}.  At present, the distance of entanglement is partly limited by the length that short wavelengths of light travel through an optical fiber, another potential advantage of Ba$^+$, since its cooling wavelength is longer than these other examples.  The ability to remotely entangle qubits has application not only for quantum repeaters~\cite{Briegel1998,Duan2004b} but also for loophole free Bell inequality tests~\cite{Simon2003}.  Necessary to the success of such an experiment is the ability to perform every operation very quickly, on the $\mu$s time scale.  In order to achieve such time scales, the power of the 1762 nm laser will have to be amplified, and the collection efficiency of the ion florescence will have to be increased.  The former task can be accomplished using a Tm based fiber amplifier presently under development.  The latter task will involve a high numerical aperture light collecting mirror placed inside the vacuum, near the trap itself.  This will greatly increase the light collection solid angle, and thereby decrease the bright/dark discrimination time, which is currently limited to a ms.  This modified trap design is presently in testing.

\section{Acknowledgements}

We would like to give special thanks to Sanghoon Chong, Tom Chartrand, Adam Kleczewski, Viki Mirgon, Joseph Pirtle and Edan Shahar for their various contributions to the experiment.  This work was supported by NSF AMO program, the ARO DURIP grant, and the University of Washington Royalty Research Fund.

\bibliography{NNP2008Proceedings,books,other}

\begin{thebibliography}{24}
\expandafter\ifx\csname natexlab\endcsname\relax\def\natexlab#1{#1}\fi
\expandafter\ifx\csname bibnamefont\endcsname\relax
  \def\bibnamefont#1{#1}\fi
\expandafter\ifx\csname bibfnamefont\endcsname\relax
  \def\bibfnamefont#1{#1}\fi
\expandafter\ifx\csname citenamefont\endcsname\relax
  \def\citenamefont#1{#1}\fi
\expandafter\ifx\csname url\endcsname\relax
  \def\url#1{\texttt{#1}}\fi
\expandafter\ifx\csname urlprefix\endcsname\relax\def\urlprefix{URL }\fi
\providecommand{\bibinfo}[2]{#2}
\providecommand{\eprint}[2][]{\url{#2}}

\bibitem[{\citenamefont{Turchette et~al.}(1998)\citenamefont{Turchette, Wood,
  King, Myatt, Leibfried, Itano, Monroe, and Wineland}}]{Turchette1998}
\bibinfo{author}{\bibfnamefont{Q.~A.} \bibnamefont{Turchette}},
  \bibinfo{author}{\bibfnamefont{C.~S.} \bibnamefont{Wood}},
  \bibinfo{author}{\bibfnamefont{B.~E.} \bibnamefont{King}},
  \bibinfo{author}{\bibfnamefont{C.~J.} \bibnamefont{Myatt}},
  \bibinfo{author}{\bibfnamefont{D.}~\bibnamefont{Leibfried}},
  \bibinfo{author}{\bibfnamefont{W.~M.} \bibnamefont{Itano}},
  \bibinfo{author}{\bibfnamefont{C.}~\bibnamefont{Monroe}}, \bibnamefont{and}
  \bibinfo{author}{\bibfnamefont{D.~J.} \bibnamefont{Wineland}},
  \bibinfo{journal}{Phys. Rev. Lett.} \textbf{\bibinfo{volume}{81}},
  \bibinfo{pages}{3631} (\bibinfo{year}{1998}).

\bibitem[{\citenamefont{Blatt et~al.}(2004)\citenamefont{Blatt, Haffner, Roos,
  Becher, and Schmidt-Kaler}}]{Blatt2004}
\bibinfo{author}{\bibfnamefont{R.}~\bibnamefont{Blatt}},
  \bibinfo{author}{\bibfnamefont{H.}~\bibnamefont{Haffner}},
  \bibinfo{author}{\bibfnamefont{C.~F.} \bibnamefont{Roos}},
  \bibinfo{author}{\bibfnamefont{C.}~\bibnamefont{Becher}}, \bibnamefont{and}
  \bibinfo{author}{\bibfnamefont{F.}~\bibnamefont{Schmidt-Kaler}},
  \bibinfo{journal}{Quant. Inf. Proc.} \textbf{\bibinfo{volume}{3}},
  \bibinfo{pages}{61} (\bibinfo{year}{2004}).

\bibitem[{\citenamefont{Lucas et~al.}(2007)\citenamefont{Lucas, Keitch, Home,
  Imreh, McDonnell, Stacey, Szwer, and Steane}}]{Lucas2007}
\bibinfo{author}{\bibfnamefont{D.~M.} \bibnamefont{Lucas}},
  \bibinfo{author}{\bibfnamefont{B.~C.} \bibnamefont{Keitch}},
  \bibinfo{author}{\bibfnamefont{J.~P.} \bibnamefont{Home}},
  \bibinfo{author}{\bibfnamefont{G.}~\bibnamefont{Imreh}},
  \bibinfo{author}{\bibfnamefont{M.~J.} \bibnamefont{McDonnell}},
  \bibinfo{author}{\bibfnamefont{D.~N.} \bibnamefont{Stacey}},
  \bibinfo{author}{\bibfnamefont{D.~J.} \bibnamefont{Szwer}}, \bibnamefont{and}
  \bibinfo{author}{\bibfnamefont{A.~M.} \bibnamefont{Steane}}
  (\bibinfo{year}{2007}), \eprint{quant-ph/0710.4421}.

\bibitem[{\citenamefont{Lee et~al.}(2003)\citenamefont{Lee, Blinov, Brickman,
  Deslauriers, Madsen, Miller, Moehring, Stick, and Monroe}}]{Lee2003}
\bibinfo{author}{\bibfnamefont{P.~J.} \bibnamefont{Lee}},
  \bibinfo{author}{\bibfnamefont{B.~B.} \bibnamefont{Blinov}},
  \bibinfo{author}{\bibfnamefont{K.}~\bibnamefont{Brickman}},
  \bibinfo{author}{\bibfnamefont{L.}~\bibnamefont{Deslauriers}},
  \bibinfo{author}{\bibfnamefont{M.~J.} \bibnamefont{Madsen}},
  \bibinfo{author}{\bibfnamefont{R.}~\bibnamefont{Miller}},
  \bibinfo{author}{\bibfnamefont{D.~L.} \bibnamefont{Moehring}},
  \bibinfo{author}{\bibfnamefont{D.}~\bibnamefont{Stick}}, \bibnamefont{and}
  \bibinfo{author}{\bibfnamefont{C.}~\bibnamefont{Monroe}},
  \bibinfo{journal}{Opt. Lett.} \textbf{\bibinfo{volume}{28}},
  \bibinfo{pages}{1582} (\bibinfo{year}{2003}).

\bibitem[{\citenamefont{Schaetz et~al.}(2007)\citenamefont{Schaetz,
  Friedenauer, Schmitz, Petersen, and Kahra}}]{Schaetz2007}
\bibinfo{author}{\bibfnamefont{T.}~\bibnamefont{Schaetz}},
  \bibinfo{author}{\bibfnamefont{A.}~\bibnamefont{Friedenauer}},
  \bibinfo{author}{\bibfnamefont{H.}~\bibnamefont{Schmitz}},
  \bibinfo{author}{\bibfnamefont{L.}~\bibnamefont{Petersen}}, \bibnamefont{and}
  \bibinfo{author}{\bibfnamefont{S.}~\bibnamefont{Kahra}}, \bibinfo{journal}{J.
  Mod. Optics} \textbf{\bibinfo{volume}{54}}, \bibinfo{pages}{2317}
  (\bibinfo{year}{2007}).

\bibitem[{\citenamefont{Letchumanan et~al.}(2007)\citenamefont{Letchumanan,
  Wilpers, Brownnutt, Gill, and Sinclair}}]{Letchumanan2007}
\bibinfo{author}{\bibfnamefont{V.}~\bibnamefont{Letchumanan}},
  \bibinfo{author}{\bibfnamefont{G.}~\bibnamefont{Wilpers}},
  \bibinfo{author}{\bibfnamefont{M.}~\bibnamefont{Brownnutt}},
  \bibinfo{author}{\bibfnamefont{P.}~\bibnamefont{Gill}}, \bibnamefont{and}
  \bibinfo{author}{\bibfnamefont{A.~G.} \bibnamefont{Sinclair}},
  \bibinfo{journal}{Phys. Rev. A} \textbf{\bibinfo{volume}{75}},
  \bibinfo{eid}{063425} (\bibinfo{year}{2007}).

\bibitem[{\citenamefont{Olmschenk et~al.}(2007)\citenamefont{Olmschenk, Younge,
  Moehring, Matsukevich, Maunz, and Monroe}}]{Olmschenk2007}
\bibinfo{author}{\bibfnamefont{S.}~\bibnamefont{Olmschenk}},
  \bibinfo{author}{\bibfnamefont{K.~C.} \bibnamefont{Younge}},
  \bibinfo{author}{\bibfnamefont{D.~L.} \bibnamefont{Moehring}},
  \bibinfo{author}{\bibfnamefont{D.}~\bibnamefont{Matsukevich}},
  \bibinfo{author}{\bibfnamefont{P.}~\bibnamefont{Maunz}}, \bibnamefont{and}
  \bibinfo{author}{\bibfnamefont{C.}~\bibnamefont{Monroe}},
  \bibinfo{journal}{Phys. Rev. A} \textbf{\bibinfo{volume}{76}},
  \bibinfo{pages}{052314} (\bibinfo{year}{2007}).

\bibitem[{\citenamefont{Blinov et~al.}(2004{\natexlab{a}})\citenamefont{Blinov,
  Leibfried, Monroe, and Wineland}}]{Blinov2004b}
\bibinfo{author}{\bibfnamefont{B.~B.} \bibnamefont{Blinov}},
  \bibinfo{author}{\bibfnamefont{D.}~\bibnamefont{Leibfried}},
  \bibinfo{author}{\bibfnamefont{C.}~\bibnamefont{Monroe}}, \bibnamefont{and}
  \bibinfo{author}{\bibfnamefont{D.~J.} \bibnamefont{Wineland}},
  \bibinfo{journal}{Quant. Inf. Proc.} \textbf{\bibinfo{volume}{3}},
  \bibinfo{pages}{45} (\bibinfo{year}{2004}{\natexlab{a}}).

\bibitem[{\citenamefont{{Neuhauser} et~al.}(1980)\citenamefont{{Neuhauser},
  {Hohenstatt}, {Toschek}, and {Dehmelt}}}]{Dehmelt1980}
\bibinfo{author}{\bibfnamefont{W.}~\bibnamefont{{Neuhauser}}},
  \bibinfo{author}{\bibfnamefont{M.}~\bibnamefont{{Hohenstatt}}},
  \bibinfo{author}{\bibfnamefont{P.~E.} \bibnamefont{{Toschek}}},
  \bibnamefont{and}
  \bibinfo{author}{\bibfnamefont{H.}~\bibnamefont{{Dehmelt}}},
  \bibinfo{journal}{Phys. Rev. A.} \textbf{\bibinfo{volume}{22}},
  \bibinfo{pages}{1137} (\bibinfo{year}{1980}).

\bibitem[{\citenamefont{DeVoe and Kurtsiefer}(2002)}]{DeVoe2002}
\bibinfo{author}{\bibfnamefont{R.~G.} \bibnamefont{DeVoe}} \bibnamefont{and}
  \bibinfo{author}{\bibfnamefont{C.}~\bibnamefont{Kurtsiefer}},
  \bibinfo{journal}{Phys. Rev. A} \textbf{\bibinfo{volume}{65}},
  \bibinfo{pages}{063407} (\bibinfo{year}{2002}).

\bibitem[{\citenamefont{Koerber et~al.}(2002)\citenamefont{Koerber, Schacht,
  Hendrickson, Nagourney, and Fortson}}]{Koerber2002}
\bibinfo{author}{\bibfnamefont{T.~W.} \bibnamefont{Koerber}},
  \bibinfo{author}{\bibfnamefont{M.~H.} \bibnamefont{Schacht}},
  \bibinfo{author}{\bibfnamefont{K.~R.~G.} \bibnamefont{Hendrickson}},
  \bibinfo{author}{\bibfnamefont{W.}~\bibnamefont{Nagourney}},
  \bibnamefont{and} \bibinfo{author}{\bibfnamefont{E.~N.}
  \bibnamefont{Fortson}}, \bibinfo{journal}{Phys. Rev. Lett.}
  \textbf{\bibinfo{volume}{88}}, \bibinfo{pages}{143002}
  (\bibinfo{year}{2002}).

\bibitem[{\citenamefont{Fortson}(1993)}]{Fortson1993}
\bibinfo{author}{\bibfnamefont{N.}~\bibnamefont{Fortson}},
  \bibinfo{journal}{Phys. Rev. Lett.} \textbf{\bibinfo{volume}{70}},
  \bibinfo{pages}{2383} (\bibinfo{year}{1993}).

\bibitem[{\citenamefont{Davidson et~al.}(1992)\citenamefont{Davidson, Snoek,
  Volten, and D\"onszelmann}}]{Davidson1991}
\bibinfo{author}{\bibfnamefont{M.~D.} \bibnamefont{Davidson}},
  \bibinfo{author}{\bibfnamefont{L.~C.} \bibnamefont{Snoek}},
  \bibinfo{author}{\bibfnamefont{H.}~\bibnamefont{Volten}}, \bibnamefont{and}
  \bibinfo{author}{\bibfnamefont{A.}~\bibnamefont{D\"onszelmann}},
  \bibinfo{journal}{Astron. Astrophys.} \textbf{\bibinfo{volume}{255}},
  \bibinfo{pages}{457} (\bibinfo{year}{1992}).

\bibitem[{\citenamefont{Silverans et~al.}(1986)\citenamefont{Silverans, Borghs,
  De~Bisschop, and Van~Hove}}]{Silverans1986}
\bibinfo{author}{\bibfnamefont{R.~E.} \bibnamefont{Silverans}},
  \bibinfo{author}{\bibfnamefont{G.}~\bibnamefont{Borghs}},
  \bibinfo{author}{\bibfnamefont{P.}~\bibnamefont{De~Bisschop}},
  \bibnamefont{and} \bibinfo{author}{\bibfnamefont{M.}~\bibnamefont{Van~Hove}},
  \bibinfo{journal}{Phys. Rev. A} \textbf{\bibinfo{volume}{33}},
  \bibinfo{pages}{2117} (\bibinfo{year}{1986}).

\bibitem[{\citenamefont{Metcalf and van~der Straten}(2001)}]{Metcalf}
\bibinfo{author}{\bibfnamefont{H.~J.} \bibnamefont{Metcalf}} \bibnamefont{and}
  \bibinfo{author}{\bibfnamefont{P.}~\bibnamefont{van~der Straten}},
  \emph{\bibinfo{title}{Laser Cooling and Trapping}}
  (\bibinfo{publisher}{Springer}, \bibinfo{year}{2001}), ISBN
  \bibinfo{isbn}{978-0-387-98728-6}.

\bibitem[{\citenamefont{Wunderlich et~al.}(2007)\citenamefont{Wunderlich,
  Hannemann, K\"orber, H\"affner, Roos, H\"ansel, Blatt, and
  Schmidt-Kaler}}]{Wunderlich2005}
\bibinfo{author}{\bibfnamefont{C.}~\bibnamefont{Wunderlich}},
  \bibinfo{author}{\bibfnamefont{T.}~\bibnamefont{Hannemann}},
  \bibinfo{author}{\bibfnamefont{T.}~\bibnamefont{K\"orber}},
  \bibinfo{author}{\bibfnamefont{H.}~\bibnamefont{H\"affner}},
  \bibinfo{author}{\bibfnamefont{C.}~\bibnamefont{Roos}},
  \bibinfo{author}{\bibfnamefont{W.}~\bibnamefont{H\"ansel}},
  \bibinfo{author}{\bibfnamefont{R.}~\bibnamefont{Blatt}}, \bibnamefont{and}
  \bibinfo{author}{\bibfnamefont{F.}~\bibnamefont{Schmidt-Kaler}},
  \bibinfo{journal}{J. Mod. Optics} \textbf{\bibinfo{volume}{54}},
  \bibinfo{pages}{1541} (\bibinfo{year}{2007}), \eprint{quant-ph/0508159}.

\bibitem[{\citenamefont{Blatt and Werth}(1982)}]{Blatt1981}
\bibinfo{author}{\bibfnamefont{R.}~\bibnamefont{Blatt}} \bibnamefont{and}
  \bibinfo{author}{\bibfnamefont{G.}~\bibnamefont{Werth}},
  \bibinfo{journal}{Phys. Rev. A} \textbf{\bibinfo{volume}{25}},
  \bibinfo{pages}{1476} (\bibinfo{year}{1982}).

\bibitem[{\citenamefont{Kurz et~al.}(2008)\citenamefont{Kurz, Dietrich, Shu,
  Bowler, Salacka, Mirgon, and Blinov}}]{Kurz2008}
\bibinfo{author}{\bibfnamefont{N.}~\bibnamefont{Kurz}},
  \bibinfo{author}{\bibfnamefont{M.~R.} \bibnamefont{Dietrich}},
  \bibinfo{author}{\bibfnamefont{G.}~\bibnamefont{Shu}},
  \bibinfo{author}{\bibfnamefont{R.}~\bibnamefont{Bowler}},
  \bibinfo{author}{\bibfnamefont{J.}~\bibnamefont{Salacka}},
  \bibinfo{author}{\bibfnamefont{V.}~\bibnamefont{Mirgon}}, \bibnamefont{and}
  \bibinfo{author}{\bibfnamefont{B.~B.} \bibnamefont{Blinov}},
  \bibinfo{journal}{Phys. Rev. A} \textbf{\bibinfo{volume}{77}},
  \bibinfo{eid}{060501} (\bibinfo{year}{2008}).

\bibitem[{\citenamefont{Blinov et~al.}(2004{\natexlab{b}})\citenamefont{Blinov,
  Moehring, Duan, and Monroe}}]{Blinov2004}
\bibinfo{author}{\bibfnamefont{B.~B.} \bibnamefont{Blinov}},
  \bibinfo{author}{\bibfnamefont{D.~L.} \bibnamefont{Moehring}},
  \bibinfo{author}{\bibfnamefont{L.-M.} \bibnamefont{Duan}}, \bibnamefont{and}
  \bibinfo{author}{\bibfnamefont{C.}~\bibnamefont{Monroe}},
  \bibinfo{journal}{Nature} \textbf{\bibinfo{volume}{428}},
  \bibinfo{pages}{153} (\bibinfo{year}{2004}{\natexlab{b}}).

\bibitem[{\citenamefont{Moehring et~al.}(2007)\citenamefont{Moehring, Maunz,
  Olmschenk, Younge, Matsukevich, Duan, and Monroe}}]{Moehring2007}
\bibinfo{author}{\bibfnamefont{D.~L.} \bibnamefont{Moehring}},
  \bibinfo{author}{\bibfnamefont{P.}~\bibnamefont{Maunz}},
  \bibinfo{author}{\bibfnamefont{S.}~\bibnamefont{Olmschenk}},
  \bibinfo{author}{\bibfnamefont{K.~C.} \bibnamefont{Younge}},
  \bibinfo{author}{\bibfnamefont{D.~N.} \bibnamefont{Matsukevich}},
  \bibinfo{author}{\bibfnamefont{L.~M.} \bibnamefont{Duan}}, \bibnamefont{and}
  \bibinfo{author}{\bibfnamefont{C.}~\bibnamefont{Monroe}},
  \bibinfo{journal}{Nature} \textbf{\bibinfo{volume}{449}}, \bibinfo{pages}{68}
  (\bibinfo{year}{2007}).

\bibitem[{\citenamefont{Garc\'ia-Ripoll
  et~al.}(2003)\citenamefont{Garc\'ia-Ripoll, Zoller, and
  Cirac}}]{GarciaRipoll2003}
\bibinfo{author}{\bibfnamefont{J.~J.} \bibnamefont{Garc\'ia-Ripoll}},
  \bibinfo{author}{\bibfnamefont{P.}~\bibnamefont{Zoller}}, \bibnamefont{and}
  \bibinfo{author}{\bibfnamefont{J.~I.} \bibnamefont{Cirac}},
  \bibinfo{journal}{Phys. Rev. Lett.} \textbf{\bibinfo{volume}{91}},
  \bibinfo{pages}{157901} (\bibinfo{year}{2003}).

\bibitem[{\citenamefont{Briegel et~al.}(1998)\citenamefont{Briegel, D\"ur,
  Cirac, and Zoller}}]{Briegel1998}
\bibinfo{author}{\bibfnamefont{H.-J.} \bibnamefont{Briegel}},
  \bibinfo{author}{\bibfnamefont{W.}~\bibnamefont{D\"ur}},
  \bibinfo{author}{\bibfnamefont{J.~I.} \bibnamefont{Cirac}}, \bibnamefont{and}
  \bibinfo{author}{\bibfnamefont{P.}~\bibnamefont{Zoller}},
  \bibinfo{journal}{Phys. Rev. Lett.} \textbf{\bibinfo{volume}{81}},
  \bibinfo{pages}{5932} (\bibinfo{year}{1998}).

\bibitem[{\citenamefont{Duan et~al.}(2004)\citenamefont{Duan, Blinov, Moehring,
  and Monroe}}]{Duan2004b}
\bibinfo{author}{\bibfnamefont{L.~M.} \bibnamefont{Duan}},
  \bibinfo{author}{\bibfnamefont{B.~B.} \bibnamefont{Blinov}},
  \bibinfo{author}{\bibfnamefont{D.~L.} \bibnamefont{Moehring}},
  \bibnamefont{and} \bibinfo{author}{\bibfnamefont{C.}~\bibnamefont{Monroe}},
  \bibinfo{journal}{Quant. Inf. and Comp.} \textbf{\bibinfo{volume}{4}},
  \bibinfo{pages}{165} (\bibinfo{year}{2004}), \eprint{quant-ph/0401020}.

\bibitem[{\citenamefont{Simon and Irvine}(2003)}]{Simon2003}
\bibinfo{author}{\bibfnamefont{C.}~\bibnamefont{Simon}} \bibnamefont{and}
  \bibinfo{author}{\bibfnamefont{W.~T.~M.} \bibnamefont{Irvine}},
  \bibinfo{journal}{Phys. Rev. Lett.} \textbf{\bibinfo{volume}{91}},
  \bibinfo{eid}{110405} (\bibinfo{year}{2003}).

\end{thebibliography}
\bibliographystyle{apsrev}

\end{document}